\documentclass{emulateapj}

\newcommand{\HA}{\ensuremath{\mathrm{H}\alpha}}

\newcommand{\OII}{[O {\sc ii}]}
\begin{document}

\title{
THE ROLE OF GALAXY INTERACTION IN ENVIRONMENTAL DEPENDENCE OF
THE STAR FORMATION ACTIVITY AT $z\simeq 1.2$}

\author{Y. Ideue\altaffilmark{1}, 
Y. Taniguchi\altaffilmark{2},
T. Nagao\altaffilmark{3,4},
Y. Shioya\altaffilmark{2},
M. Kajisawa\altaffilmark{2},
J. R. Trump\altaffilmark{5},
D. Vergani\altaffilmark{6},
A. Iovino\altaffilmark{7},
A. M. Koekemoer\altaffilmark{8},
O. Le F\`evre\altaffilmark{9},
O. Ilbert\altaffilmark{10}, and
N. Z. Scoville\altaffilmark{11}
}

\altaffiltext{*}{Based on observations with the NASA/ESA 
        {\it Hubble Space Telescope}, obtained at the Space Telescope Science 
	Institute, which is operated by AURA Inc, under NASA contract NAS 
	5-26555. Also based on observations made with the Spitzer Space 
	Telescope, which is operated by the 
	Jet Propulsion Laboratory, California Institute of Technology, 
	under NASA contract 1407. Also based on data collected at;  
	the Subaru Telescope, which is operated by the National Astronomical 
	Observatory of Japan; the XMM-Newton, an ESA science mission with 
	instruments and contributions directly funded by ESA Member States and
	NASA; the European Southern Observatory under Large 
	Program 175.A-0839, Chile; Kitt Peak National Observatory, Cerro 
	Tololo Inter-American Observatory and the National Optical Astronomy 
	Observatory, which are operated by the Association of Universities for 
	Research in Astronomy, Inc. (AURA) under cooperative agreement with 
	the National Science Foundation; and the Canada-France-Hawaii 
	Telescope with MegaPrime/MegaCam operated as a joint project by the 
	CFHT Corporation, CEA/DAPNIA, the NRC and CADC of Canada, the CNRS 
	of France, TERAPIX and the Univ. of Hawaii.}

\altaffiltext{1}{Graduate School of Science and Engineering, Ehime University, 
        Bunkyo-cho, Matsuyama 790-8577, Japan:
      {\it e-mail: ideue@cosmos.phys.sci.ehime-u.ac.jp}}
\altaffiltext{2}{Research Center for Space and Cosmic Evolution, 
        Ehime University, Bunkyo-cho, Matsuyama 790-8577, Japan}
\altaffiltext{3}{The Hakubi Project, Kyoto University, Yoshida-Ushinomiya-cho,
Sakyo-ku, Kyoto 606-8302, Japan}
\altaffiltext{4}{Department of Astronomy, Kyoto University, Kitashirakawa-Oiwake-cho,
Sakyo-ku, Kyoto 606-8502, Japan}
\altaffiltext{5}{UCO/Lick, UC Santa Cruz, Santa Cruz, CA 95064 USA}
\altaffiltext{6}{INAF-Osservatorio Astronomico di Bolona, Via Ranzani 1, I-40127, Bologna, Italy}
\altaffiltext{7}{INSF-Osservatorio Astronomico di Brera, Via Brera 28, I-20159 Milano, Italy}
\altaffiltext{8}{Space Telescope Science Institute, 3700 San Martin Drive, Baltimore, MD 21218}
\altaffiltext{9}{Laboratoire d'Astrophysique de Marseile, CNRS-Universit\'e d'Aix-Marseille, 38 rue Frederic Joliot Curie, F-13388 Marseille, France}
\altaffiltext{10}{Observatoriore de Marseille-Provence, Pole de I'Etoile Site
de Chiateau-Gombert, 38 rue Frederic Joliot-Curie, 13388 Marseille cedex 13, 
France}

\altaffiltext{11}{Department of Astronomy, MS 105-24, California Institute of
                Technology, Pasadena, CA 91125}

\shortauthors{Ideue et al.}
\shorttitle{Relation between star formation on stellar mass \& environments at $z \simeq 1.2$}

%==============================================================================
% ABSTRACT
%==============================================================================
\begin{abstract}

In order to understand environmental effects on star formation in
high-redshift galaxies, we investigate the physical relationships
between the star formation activity, stellar mass, and environment for
$z\simeq1.2$ galaxies in the 2 deg$^2$ COSMOS field$^*$.  We estimate star
formation using the \OII$\lambda 3727$ emission line and environment
from the local galaxy density.  Our analysis shows that for massive
galaxies ($M_* \gtrsim 10^{10} M_{\sun}$), the fraction of \OII\ emitters in
high-density environments ($\Sigma_{\rm 10th}\gtrsim 3.9\ {\rm
  Mpc^{-2}}$) is $1.7\pm 0.4$ times higher than in low-density
environments ($\Sigma_{\rm 10th} \lesssim 1.5\ {\rm Mpc^{-2}}$), while
the \OII\ emitter fraction does not depend on environment for low-mass
$M_* \lesssim 10^{10} M_{\sun}$ galaxies.  In order to understand
what drives these trends, we investigate the role of companion
galaxies in our sample.  We find that the fraction of \OII\ emitters
in galaxies with companions is $2.4\pm0.5$ times 
as high as that in galaxies without
companions at $M_*\gtrsim 10^{10} M_{\sun}$.  In
addition, massive galaxies are more likely to have companions in
high-density environments.  However, although the {\it number} of star
forming galaxies increases for massive galaxies with close companions
and in dense environments, the {\it average} star formation rate
of star forming galaxies at a
given mass is independent of environment and the presence/absence of a
close companion.  These results suggest that interactions and/or
mergers in high-density environment could induce star formation in
massive galaxies at $z\sim1.2$, increasing the fraction of
star-forming galaxies with $M_* \gtrsim 10^{10} M_{\sun}$.

\end{abstract}


%\maketitle

%==============================================================================
% 1.INTRODUCTION
%==============================================================================
\section{INTRODUCTION}
 
A key question in understanding the formation and evolution of
galaxies is to find what physical parameters are most sensitive to the
star formation process in galaxies; e.g., how the star formation
activity depends on the environment, and how the relation between star
formation and environment changes during the course of galaxy
evolution over 10 Gyrs.  Observational properties of galaxies in the
local universe have been extensively studied in the past.  Typically,
actively star-forming galaxies in the local universe have lower masses
than passive galaxies and the most massive galaxies tend to be
inactive for star formation (so-called mass-downsizing; Cowie et
al. 1996).  In addition, the star formation activity strongly depends
on environment: star formation rate (SFR) decreases with increasing
galaxy density (e.g., Lewis et al. 2002; Gomez et al.2003; Kauffmann
et al. 2004; Mahajan et al. 2010) and the fraction of star-forming
galaxies also decreases with increasing galaxy density (e.g., Carter
et al. 2001; Balogh et al. 2004; Mateus \& Sodr\'e 2004).  The
fraction of early-type (passive) galaxies is higher in higher-density
regions while the fraction of late-type (star-forming) ones is lower
in such environment, called the ``morphology-density relation'' and
consistent with the above results (e.g., Dressler 1980; Dressler et
al. 1997; Goto et al. 2003; Capak et al. 2007).  These findings
indicate that the star formation activity in galaxies is strongly
related to both the stellar mass and their environment.

On the other hand, it is well known that the star formation rate
density (SFRD) steeply increases in the first $\sim 2$ Gyrs, peaks at
$z\sim 1-3$, and then decreases by an order of magnitude toward the
present day (e.g., Madau et al.1996; Lilly et al. 1996; Shioya et
al. 2008; Bouwens et al. 2009).  These results suggest that the
redshift range of $z \sim 1-3$ is the key epoch for the most active
cosmic star formation in the history of universe.  Therefore, studies
of galaxies at $1 < z < 3$ are important for understanding galaxy
evolution.  In recent years, wide and deep surveys have allowed us to
study star formation activity in $z \gtrsim 1$ galaxies.  For example,
Elbaz et al. (2007) and Cooper et al. (2008) studied the relation
between the star formation rate (SFR) in galaxies and the environment
at $z\sim 1$, in the Great Observatories Origins Deep Survey (GOODS;
Giavalisco et al. 2004) and the DEEP2 Galaxy Redshift Survey (DEEP2;
Davis et al. 2003).  These studies showed that the average star
formation rate increases with increasing galaxy density (Elbaz et
al. 2007; Cooper et al. 2008).  We also investigated the relation
between the fraction of star-forming galaxies with \OII\ $\lambda
3727$ emission (hereafter ``\OII\ emitters'') and the local galaxy
density at $z\simeq 1.2$ in the Cosmic Evolution Survey (COSMOS;
Scoville et al. 2007a; Koekemoer et al. 2007) and found that the
fraction of star-forming galaxies in high-density regions is as high
as that in low-density regions (Ideue et al. 2009; hereafter ``Paper
{\sc i}'').  Similar trends were also reported by other observational
studies in groups of galaxies (Iovino et al. 2010; Li et al. 2011;
Sobral et al. 2011) and cluster environments (Hayashi et al. 2010;
Hilton et al. 2010) at $z \gtrsim 1$.  Since the fraction of
star-forming galaxies decreases with increasing galaxy density in the
local universe as mentioned above, these results suggest that the
relationship between the star formation activity in galaxies and their
environment dramatically changed from $z\sim1$ to the present day.  On
the other hand, many studies suggest that the star formation activity
or the fraction of star-forming galaxies also strongly depends on
stellar mass even at $z\gtrsim 1$ (e.g., Noeske et al. 2007; Elbaz et
al. 2007; Ilbert et al. 2010).  Therefore, it is important to
investigate the relationships between the star formation activity, the
stellar mass of galaxies, and the environment at $z\sim 1$, in order
to reveal the origin of the observed difference in the environmental
dependence of the star formation activity between $z\sim0$ and
$z\sim1$.

In this paper, we focus on the relationships between the star
formation, stellar mass, and environment of galaxies at $z \simeq 1.2$
in the COSMOS field. We have already carried out Subaru imaging
observations of the COSMOS field (Taniguchi et al. 2007). We can
obtain the sample of \OII\ emitters in the COSMOS field by using a
narrowband filter NB816 ($\lambda_c = 8150 $ \AA\ and $\Delta \lambda
{\rm (FWHM)} = 120 $ \AA, see Takahashi et al. 2007) and estimate
their SFR by using the \OII\ luminosity.  The \OII\ emission line
provides a good estimator of the star formation in galaxies at
intermediate redshift (e.g., Kennicutt 1998; Jansen et al. 2001).  The
stellar mass in galaxies in the COSMOS field was obtained from the
fitting of their spectral energy distributions (SEDs; Ilbert et
al. 2010).  Thanks to the very large survey area ($\sim$ 2 deg$^2$) of
the COSMOS field, we are able to obtain an unbiased picture of the
star formation activity in galaxies, avoiding the cosmic variance
effect.  Indeed, the COSMOS field includes various regions with a wide
range of galaxy density (e.g., Scoville et al. 2007a, 2007b; Takahashi
et al. 2007), which enables us to investigate systematically the
environmental effects on galaxy properties.

This paper is organized as follows. In section 2, we describe the
sample selection and possible contamination of the \OII\ emitter
sample.  Descriptions of the measurements of local galaxy density,
stellar mass, and SFR are presented in section 3.  We investigate the
dependence of star formation on environment and stellar mass, and the
effect of close companions on the star formation activity, for our
sample at $z\simeq 1.2$ in section 4.  In section 5, we summarize our
findings and discuss implications of our results on the evolution of
galaxies from $z\sim 1.2$ to $z\sim 0$.  Throughout this paper,
magnitudes are given in the AB system. We adopt a flat universe with
the following cosmological parameters; $\Omega_{\rm matter} = 0.3$,
$\Omega_{\Lambda} = 0.7$, and $H_0 = 70\; {\rm km\;s^{-1}\;Mpc^{-1}}$.

%==============================================================================
% 2. THE SAMPLE
%==============================================================================
\section{THE SAMPLE}

%------------------------------------------------------------------------------
% 2.1 catalog
%------------------------------------------------------------------------------
\subsection{Catalog}

The COSMOS field has multi-wavelength photometric data over an area of
$\sim$2 deg$^2$ from X-ray to radio.  We use the COSMOS intermediate-
and broad-band photometry catalog which includes 16 broad, 12
intermediate, and 2 narrow bands (Capak et al. 2011, in preparation:
hereafter ``the COSMOS photometric catalog'').  We also use the
photometric redshift catalog (Ilbert et al. 2009: hereafter ``the
COSMOS photo-$z$ catalog'') which includes 937013 objects whose total
$i$ magnitudes (Subaru $i^\prime$ or CFHT $i^*$) are brighter than 26
mag.  Photometric redshifts in COSMOS are computed using the 25
optical bands, plus 2 GALEX and 4 Spitzer/IRAC bands (Ilbert et
al. 2009, see also Section 3.2).

%------------------------------------------------------------------------------
% 2.2 Sample Selection
%------------------------------------------------------------------------------
\subsection{Sample Selection}

In order to select \OII\ emitters at $z\sim1.2$, we use the
photometric data of $i^\prime$, $z^\prime$, and $NB816$ bands taken on
the Subaru Telescope and those of $i^*$-band taken on the Canada
France Hawaii Telescope in the COSMOS photometric catalog (Capak et
al. 2011, in preparation). We also use the photometric redshifts from
the COSMOS photo-$z$ catalog (Ilbert et al. 2009).

Using the $NB816$-band photometric data, Takahashi et al. (2007)
extracted a sample of \OII\ emitters at $1.17 < z < 1.20$.  Here we
additionally extract a sample of non-\OII-detected galaxies in the
same $1.17 < z_{\rm ph} < 1.20$ redshift range.  Since fainter
galaxies have larger photo-$z$ errors, we select only galaxies with
$i^{\prime} < 24$, for which the photo-$z$ error is $\sigma_{\Delta z}
= 0.026$ at $z \sim 1.2$.  Consequently, we obtain 1654 $i$-selected
galaxies with $1.17< z_{\rm ph} < 1.20$.
 
Among the 1654 galaxies, we select galaxies satisfying the following
criteria as \OII\ emitters:
\begin{equation}
iz-NB816 > \max (0.2,3\sigma_{iz-NB816}), 
\end{equation}
where 
\begin{equation}
3\sigma_{iz-NB816}=-2.5\log (1-\sqrt{f_{3\sigma_NB816}^2 + f_{3\sigma_{iz}}^2}/f_{NB816}), 
\end{equation}
and 
\begin{equation}
f_{iz} = 0.57f_i + 0.43f_z. 
\end{equation}
Note that $iz-NB816=0.2$ corresponds to $EW(\textrm{\OII})
\approx 12$ \AA\ in the rest frame, where $iz$ is the matched
continuum for the central wavelength of NB816 filter (see Takahashi et
al. 2007).

We obtain 932 \OII\ emitters among the photo-$z$ selected galaxies
with $i^\prime < 24$.  The remaining 722 objects are galaxies without
a significant narrow-band excess by the \OII\ emission line; their
rest-frame \OII\ emission equivalent width is $EW (\textrm{\OII}) <
12$ \AA\ and we designate them ``non-\OII\ emitters'' in this paper.
%A summary of the samples is given in Table \ref{tab:sample}. 

%----------------------------------------------------------------------------
% 2.3 Contamination in the OII emitter sample
%----------------------------------------------------------------------------
\subsection{Contamination and Incompleteness due to Photometric Redshift Errors}

As mentioned in the previous section, we selected our photo-$z$ sample
and \OII\ emitters at $z\simeq 1.2$ using the photometric redshift and
eq (1).  Although the contamination is thought to be small due to the
fairly accurate photo-$z$ (Ilbert et al. 2009), we examine how our
photo-$z$ samples are contaminated by foreground or background
objects, using spectroscopic redshifts (spec-$z$) from the zCOSMOS
redshift catalog (Lilly et al. 2007; Lilly et al. in prep).  The
spec-$z$ is available for 48 \OII\ emitters and 37 non-\OII\ emitters
in our photo-$z$ sample.  Among these objects, 44 \OII\ emitters and
24 non-\OII\ emitters have spec-$z$ within our target redshift range,
$z=1.17-1.20$, while the remaining objects have $z_{\rm sp} < 1.17$ or
$z_{\rm sp} > 1.20$.  This means that the contamination rates for the
\OII\ emitter and non-\OII\ emitter samples are $8\%$ (4/48) and
$35\%$ (13/37), respectively.  Thus, the contamination rate for our
photo-$z$ sample (932 \OII\ emitters $+$ 722 non-\OII\ emitters) is
expected to be $\sim 20\%$.

We also estimate the incompleteness of our photo-$z$ sample, using 90
objects with $i^\prime < 24$ and $z_{\rm sp} = 1.17-1.20$.  Only 69
objects among them have $z_{\rm ph} = 1.17-1.20$, and so the
incompleteness is $23\%$ (21/90).  In order to investigate how
incomplete our \OII\ emitter and non-\OII\ emitter samples are, we
divided the 90 objects with $i^\prime < 24$ and $z_{\rm sp} =
1.17-1.20$ into the narrow-band excess objects and the others, using
eq. (1).  As a result, 47 objects are narrow-band excess objects, and
the other 43 objects do not show a significant narrow-band excess.
All of the 47 narrow-band excess objects with $z_{\rm sp} = 1.17-1.20$
have $z_{\rm ph} = 1.17-1.20$ and are selected as \OII\ emitters,
suggesting that photometric redshifts are nearly complete for the
\OII\ emitters.  On the other hand, 22 out of 43 objects without a
significant narrow-band excess have $z_{\rm ph} = 1.17-1.20$.  Thus
the incompleteness rates for the \OII\ emitter and non-\OII\ emitter
samples are estimated to be $0\%$ (0/47) and $49\%$ (21/43),
respectively.

In summary, the contamination and incompleteness rates for our
photo-$z$ sample are $\sim 20\%$.  To keep the uniformity of photo-$z$
selected sample, we neither exclude the contamination nor add the
missed objects into our sample, using the spectroscopic information.
We will discuss how the contamination and incompleteness due to the
photometric redshift errors for our photo-$z$ and \OII\ emitter
samples affect our results in Section 4.

%----------------------------------------------------------------------------
% 2.4 Contamination of AGNs
%----------------------------------------------------------------------------
\subsection{AGN Contamination in \OII\ Emitter Sample}

Since active galactic nuclei (AGNs) also show emission lines including
\OII\ in general, we have to consider the contribution of such AGNs in
our analysis.  Here we try to estimate possible AGN contamination in
our \OII\ emitter sample by taking account of the difference of SEDs
between AGNs and star-forming galaxies.

Stern et al. (2005) found that AGNs can be distinguished from
star-forming galaxies using a $[3.6]-[4.5]$ vs. $[5.8]-[8.0]$
color-color diagram.  While the ultraviolet to mid-infrared ($\lambda
< 5 \mu $m) continuum of star-forming galaxies is the composite
stellar black body, an AGN continuum is well fit by a power law.
Accordingly the infrared SED of AGNs tends to be systematically redder
than star-forming galaxies.  Although Stern et al. (2005) defined the
AGN selection criteria for objects at $z=0-4$, some star-forming
galaxies at $z\sim 1.2$ could be misidentified as AGNs by their
criteria.  In order to avoid this situation, we have newly defined our
own AGN criteria for objects at $z\simeq 1.2$ examining Figures 2 and
3 in Stern et al. (2005).  Our criteria are $([3.6]-[4.5])_{\rm AB} >
0$ and $([5.8]-[8.0])_{\rm AB}>0$, which are derived from the four
IRAC bands (3.6, 4.5, 5.8, and 8.0 $\mu$m) in the COSMOS photometric
catalog.
 
We apply these criteria for the 587 \OII\ emitters detected in all the
IRAC bands, and then find that 21 objects are identified as AGN
candidates.  Of the 345 objects without detection in all four IRAC
bands, 244 are detected in both 3.6 and 4.5 $\mu$m.  For these we are
able to examine AGN candidates only by using the $([3.6]-[4.5])_{\rm
  AB}$ color and $[3.6]_{\rm AB}-[4.5]_{\rm AB} > 0$; see Figure 1 in
Stern et al. (2005).  We find find that 18 of the 244 galaxies are
identified as AGNs by this criterion.  Thus, the total number of the
AGN candidates in our \OII\ emitter sample is 39, giving the possible
AGN contamination rate of 4.7\%.

It is noted, however, that our mid-infrared color criteria described
above may not be useful in identification of LINERs that are also
popular \OII\ emitters in the local universe.  For such cases, we are
able to use optical colors to estimate the contamination from LINERs
since host galaxies of LINERs tend to show red optical colors.  For
example, Yan et al. (2006) found that most \OII\ emitters with red
optical colors tend to have the LINER activity rather than star
formation.  Lemaux et al. (2010) also found that most of
\OII\ emitters on the so-called red sequence at $z=0.8-0.9$ are
identified as either LINERs or Seyferts.  In order to examine the
contamination from LINERs in our \OII\ emitter sample, we investigate
the SED of the \OII\ emitters and found that almost all galaxies show
blue colors, being comparable to those of starburst or local Sd --
Sdm-type galaxies.  Therefore, we conclude that the contamination from
LINERs is negligibly small in our sample.

Thus the possible fraction of AGN in our \OII\ emitters is $\sim 5$\%.
This is roughly consistent with the result of Garn et al. (2010) who
studied the star formation activity in galaxies at $z=0.84$ using a
sample of \HA\ emitters and found that the AGN fraction is 5-11\%.

Because the AGN contamination is negligibly small, we do not
exclude the AGN candidates from the \OII\ emitter sample in the
following analysis.  However we additionally performed the same
analysis given in Section 4 for the sample without the AGN candidates
and found the same results, confirming that the contamination from
AGNs is negligibly small.

%============================================================================== 
% 3. Description of the measurement of the local galaxy density and sSFR
%==============================================================================
\section{MEASUREMENT OF LOCAL DENSITY, STELLAR MASS, SFR}
%------------------------------------------------------------------------------
%3.1 Measurement of the local galaxy density
%------------------------------------------------------------------------------
\subsection{Local Galaxy Density}

In order to investigate the relation between the star formation
activity in galaxies and their environment, we quantify the
environment from the observational data.  We adopt the projected
$N$th-nearest-neighbor surface density since this density has been
often used as an indicator of the galaxy environment in a number of
previous investigators (e.g., Dressler 1980; Postman et al. 2005).  It
is also noted that such projected density measurements based on
photo-$z$ are useful in detecting the so-called large-scale structures
and reconstructing the overdensity (Scoville et al. 2007b).

To be consistent with previous studies (e.g., Dressler 1980; Capak et
al. 2007; Feruglio et al. 2010), we use the 10th nearest neighbor in
our analysis.  The local galaxy density is calculated by using the
projected proper distance to the 10th nearest neighbor ($r_{\rm 10th}$
in Mpc) as
\begin{equation}
\Sigma_{\rm 10th} = \frac{11}{\pi r_{\rm 10th}^2}. 
\end{equation}
We compute the projected densities for our sample galaxies using the
COSMOS photo-$z$ catalog.  We use a redshift slice centered on each
sample galaxy with a width of $\pm \sigma_{\Delta z}$ ($\sigma_{\Delta
  z}=0.026$ for $z \sim 1.2$; Ilbert et al. 2009).  Note that we use
only galaxies with $i'<24$ in a given redshift slice to secure the
photo-$z$ accuracy in the estimate of the projected densities.  For
204 galaxies near the edge of our survey field, the $r_{\rm10th}$ is
greater than the distance from the object to the field edge.  We do
not use them in the later analysis, since we cannot estimate their
local galaxy densities accurately.

\begin{figure}
\begin{center}
	\includegraphics[width=70mm]{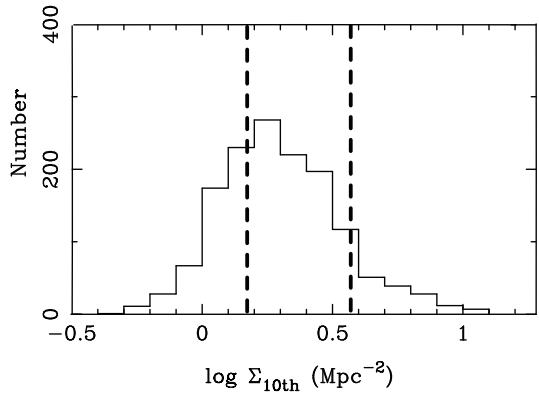} 
\caption{
The histograms show the numbers of galaxies with $i^\prime < 24$ and 
$z_{\rm ph} = 1.17-1.20$ in the COSMOS field 
 as a function of the local galaxy density derived from the distance to the 10th nearest 
neighbor. 
The vertical dashed lines show the boundaries 
between three environments.
}
\label{density-dist}
\end{center}
\end{figure}

In Figure \ref{density-dist}, we show the distribution of the local
galaxy density for our photo-$z$ selected sample.  The local galaxy
density for our sample galaxies ranges over $-0.3 \lesssim \log
\Sigma_{\rm 10th} \lesssim 1.1$.  In order to investigate
environmental effects on star formation in section 4, we define three
environment bins: low-density ($\log \Sigma_{\rm 10th} \leq 0.185$),
intermediate ($0.185 < \log \Sigma_{\rm 10th} < 0.585$), and
high-density ($0.585 \leq \log \Sigma_{\rm 10th} $).  The number of
galaxies in each environment bin is 464, 820, and 162 for the low-,
intermediate-, and high-density environment, respectively.  We note
that the intermediate-density environment includes the mean local
galaxy density in our sample ($<\log \Sigma_{\rm 10th}>=0.3$).

%--------------------------
% fig: Distribution density
%--------------------------

%------------------------------------------------------------------------------
%3.2 Measurement of stellar masses and star formation rates
%------------------------------------------------------------------------------
\subsection{Stellar Mass and Star Formation Rate}

In this study, we use stellar mass estimates which come from SED
fitting by Ilbert et al. (2010).  The COSMOS 31 photometric-band data
(including the IRAC mid-infrared data) are fit with stellar population
synthesis models (Bruzual and Charlot 2003), assuming a Chabrier
(2003) IMF and exponentially declining star formation histories (see
Ilbert et al. 2010 for details).

To estimate the SFR, we use the \OII$\lambda 3727$ emission-line
luminosity.  We measure the \OII\ luminosity from the line flux which
is estimated by using the flux densities in $i'$, $z'$, and $NB816$ as
adopted in Takahashi et al. (2007), given by
\begin{equation}
f_{\rm line}  = \Delta NB816 \frac {f_{NB816} - f_{iz}}{1-0.57(\Delta NB816/\Delta i)},
\end{equation}
where $\Delta NB816$ and $\Delta i$ are the effective bandwidth of the
NB816 and $i^\prime$ filters, respectively.  The \OII\ luminosity is
estimated from the line flux by $L(\textrm{\OII}) = 4\pi d^{2}_{\rm
  L}f_{\rm cor}(\textrm \OII)$.  In this procedure, we assume that all
the \OII $\lambda 3727$ emitters are located at $z=1.187$, the
redshift corresponding to the central wavelength of NB816 filter,
setting the luminosity distance as $d_{\rm L}=8190$ Mpc.

We then apply an extinction correction to the \OII\ luminosity of each
object.  Previous studies have generally assumed $A_{\rm
  H\alpha}=1.0$, which is estimated from observations of local
star-forming galaxies (Kennicutt 1983; Kennicutt 1998; Hopkins 2004).
However, Gilbank et al (2010; hereafter G10) found that there is a
stellar-mass dependence of the dust extinction of nebula emission
lines, with $A_{\rm H\alpha}=1.0$ overestimating the line luminosity
for low-mass galaxies and underestimating it for high-mass galaxies.
This effect should be taken into account in the estimate of SFR, since
we investigate the dependence of SFR on stellar mass in this study.
We calculate $A_{\rm H\alpha}$ for each \OII\ emitter using the
relation between $A_{\rm H\alpha}$ and stellar mass obtained by G10:
\begin{equation}
A_{\rm H\alpha}=-51.201-11.199\log M_*+0.615(\log M_*)^2
(M_* > 10^{9} M_{\sun})
\end{equation}
\begin{equation}
A_{\rm H\alpha}=0.225 
(M_*<10^{9} M_{\sun}). 
\end{equation}
Note, however, that we adopt
\begin{equation}
A_{\rm H\alpha}=1.101\log M_*-10.299
(M_*>10^{10} M_{\sun}). 
\end{equation}
because $A_{\rm H\alpha}$ estimated from eq. (9) in G10 seems to
overestimate at $M_*>10^{10} M_{\sun}$ (see Figure 4 in
G10).  We derive $A_{\rm [OII]}$ from $A_{\rm H\alpha}$ using the
Galactic extinction curve (Cardelli et al. 1989).  The
extinction-corrected \OII\ luminosity is given by
\begin{equation}
L_{\rm cor}(\textrm{\OII}) = L_{\rm line} \times 10^{0.4A_{\rm [OII]}}.
\end{equation}
We estimate the SFR from the \OII\ luminosity using the Kennicutt
(1998) relation:
\begin{equation}
{\rm SFR\ ({\it M}_{\sun}\ yr^{-1})}= 1.41 \times 10^{-41} L(\textrm{\OII}) {\rm (ergs \ s^{-1})}. 
\end{equation}

In the following analysis, we use 1446 galaxies (824 \OII\ emitters
and 622 non-\OII\ emitters) with $M_* > 10^8 M_{\sun}$ for
which the local galaxy density is available.  A summary of all galaxy
samples used in our analysis is given in Table \ref{tab:subsample}.

%--------------------
% table sample (emitter/non-emitter)
%---------
%\clearpage

\begin{deluxetable*}{rcccrcccrccc}
%\rotate
\tablecolumns{12}
\tablewidth{0pc}
\tablecaption{\label{tab:subsample} The number of galaxies in each 
environments and mass bins}
\tablehead{
\colhead{}   &  \multicolumn{3}{c}{All samples} & 
\colhead{}   &  \multicolumn{3}{c}{[O {\sc ii}] emitters} &  
\colhead{}   &  \multicolumn{3}{c}{non-[O {\sc ii}] emitters}  \\
\cline{2-4} \cline{6-8} \cline{10-12} \\
\colhead{$\log M_* (M_{\sun})$} & 
\colhead{Low} & \colhead{Intermediate} & \colhead{High} & \colhead{}    & 
\colhead{Low}   & \colhead{Intermediate}    & \colhead{High}  & \colhead{} &
\colhead{Low}   & \colhead{Intermediate}    & \colhead{High}}
\startdata
8 -- 8.5  & 0 & 4 & 0 &
          & 0 & 3 & 0 &
          & 0 & 1 & 0 \\
8.5 -- 9  & 19 & 30 & 9&
          & 13 & 20 & 8&
          & 6 & 10  & 1 \\
9 -- 9.5  & 103 & 172 & 24 &
          & 97 & 155 & 23&
          & 6 & 17 & 1 \\
9.5 -- 10 & 147 & 281 & 50 &
          & 110 & 228 & 39&
          & 37 & 53 & 11 \\
10 -- 10.5 & 82 & 139 & 48 &
           & 33 & 55 & 23 &
           & 49 & 84 & 25\\
10.5 -- 11 & 80 & 141 & 24 &
           & 5 & 5 & 4&
           & 75 & 136 & 20  \\
11 -- 11.5 & 32 & 51 & 7 &
           & 1 & 2 & 0&
           & 31 & 49 & 7\\
11.5 -- 12 & 1 & 2 & 0 &
           & 0 & 0 & 0 &
           & 1 & 2 & 0 \\
\hline
Total &464 & 820 &  162 &
      &259 &468 & 97 &
      &205 &352 &65 \\
\enddata
\end{deluxetable*}

%==============================================================================
% 4. RESULTS
%==============================================================================
\section{RESULTS}
%------------------------------------------------------------------------------
% 4.1 The relation between emitter fraction, stellar mass, and environment
%------------------------------------------------------------------------------
\subsection{The relations between star formation, stellar mass, and environment}

\begin{figure}
\begin{center}
	\includegraphics[width=75mm]{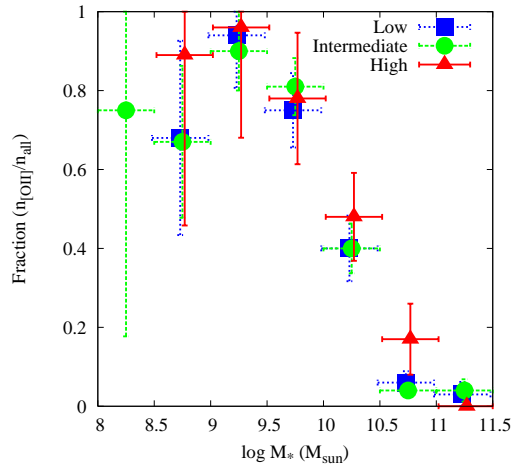} 
\caption{
The fraction of \OII\ emitters in the photo-$z$ selected sample 
as a function of stellar mass in the three different environments.
The blue squares, green circles, and red triangles represent the
low-, intermediate-, and high-density environments, respectively.
The error bars along the vertical axis are 1 $\sigma$ Poisson errors, and
those along the x-axis indicate the bin widths.
Data points for the different environments are plotted with small
horizontal offsets for clarity.}
\label{fracoii-mass-env}
\end{center}
\end{figure}

In this section, we study the relations between star formation,
stellar mass, and environment.  In Figure \ref{fracoii-mass-env}, we
show the fraction of the \OII\ emitters in the photo-$z$ selected
sample (hereafter, ``\OII\ fraction'') as a function of stellar mass
for the three different environments.  In all environments, the
\OII\ fraction decreases with increasing stellar mass, indicating that
most low-mass galaxies are star-forming while non-star-forming
galaxies dominate at high mass.  This suggests that mass-downsizing is
already in place at $z\simeq1.2$.  Several studies have reported
similar results for galaxies at $z\sim0.8$ (Iovino et al. 2010; Sobral
et al. 2011).

Next we focus on the environmental dependence of the \OII\ fraction.
While there is no environmental dependence of the \OII\ fraction in
low-mass bins ($M_* \lesssim 10^{10} M_{\sun}$), the
\OII\ fraction in the high-density environment is a little higher than
that in the intermediate- and low-density environments in high-mass
bins ($M_* \sim 10^{10}-10^{11} M_{\sun}$), although the
uncertainty for the high-density environment is relatively large due
to the small number statistics.  
We estimate the \OII\ fraction in the high- and low-density environment
for massive galaxies ($M_* \gtrsim 10^{10} M_{\sun}$).
We then find the \OII\ fraction in the high-density environment is
$1.7\pm0.4$ times higher than that in the low-density environment.
This may indicate that the star
formation activity in relatively massive galaxies is enhanced in
high-density regions.

%-----------------------------
% fig: emitter fraction  vs. M*
%-----------------------------

%------------------------------
% fig: distribution of stellar mass in \OII emitters
%------------------------------
\begin{figure}
\begin{center}
	\includegraphics[width=85mm]{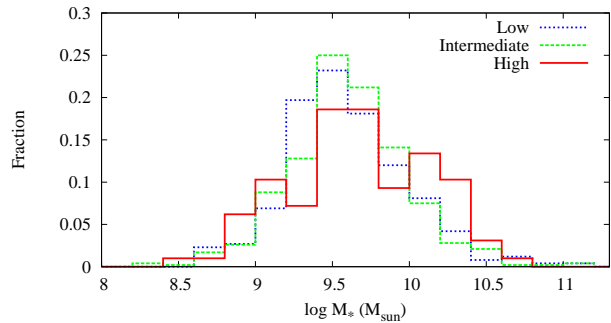} 
\caption{
The normalized histograms of stellar mass for the \OII\ emitter sample.
The blue dotted, green dashed, blue solid lines represent the 
low-, intermediate-, high-density environments, respectively.}
\label{mass-his-oii}
\end{center}
\end{figure}

%---------------------------------
% fig: sfr vs. stellar mass (environment)
%---------------------------------
\begin{figure}
\begin{center}
	\includegraphics[width=85mm]{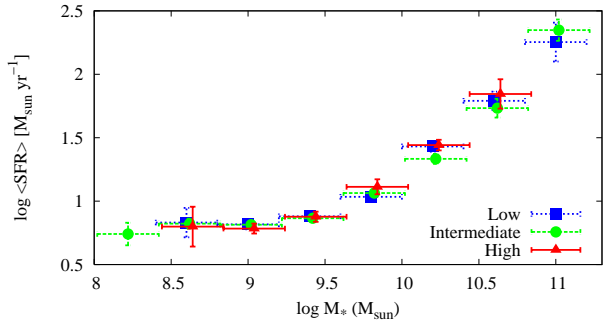} 
\caption{
The average SFR of \OII\ emitters as a function of stellar mass in 
the three different environment.
Symbols are the same as Figure \ref{fracoii-mass-env}. 
The error bars along the y-axis are 1 $\sigma$ errors based on the standard deviation,
and those along the x-axis indicate the size of bins.
Data points for the different environments are plotted with small
horizontal offsets for clarity.}
\label{sfr-mass-env}
\end{center}
\end{figure}

Figure \ref{mass-his-oii} shows the distribution of stellar mass for
the \OII\ emitters in the three different environments.  The
distribution for \OII\ emitters in the high-density environment
appears to be slightly skewed toward high mass, compared with those
for the lower-density environments.  We estimate the fraction of
high-mass galaxies ($M_* > 10^{10} M_{\sun}$) in the \OII\ emitter
sample for the low-, intermediate-, and high-density environments are
15\%, 13\%, and 28\%, respectively.  Indeed, the fraction of high-mass
\OII\ emitters in the high-density environment is more than twice as
high as the low-, and intermediate-density environments.  We apply the
Kolmogorov-Smirnov (K-S) test to the distribution of stellar mass for
the low- and high-density environments.  The K-S test gives a $1.5\%$
probability of the two distributions being drawn from the same parent
distribution.  This seems to be consistent with the above result: the
fraction of \OII\ emitters with $M_* > 10^{10} M_{\sun}$ is highest
in the high-density environment because the shape of the stellar mass
function for all of our photo-$z$ selected galaxies is similar among
the different environments over $M_* \gtrsim 10^{9} M_{\sun}$.

We investigate the average SFR as a function of stellar mass for
\OII\ emitters in the three environments (Figure \ref{sfr-mass-env}).
The linear fit gives the following relations:
$\log {\rm <SFR>}=(0.60\pm0.11)\times\log M_{*}-(4.71\pm1.09)$ 
for the low-density environment,
$(0.54\pm0.12)\times\log M_{*}-(4.20\pm1.18)$ 
for the intermediate-density environment, and
$(0.59\pm0.08)\times\log M_{*}-(4.56\pm0.78)$ 
for the high-density environment.
These results show the average SFR strongly correlates with the stellar
mass, consistent with the results in previous studies (e.g., Elbaz et
al. 2007; Daddi et al. 2007).  We also find that the relation between
SFR and stellar mass in our star-forming galaxies does not depend on
environment.  These results imply that there is no environmental
effect on the {\it strength} of star formation activity in
star-forming galaxies, although environment does affect the {\it
  fraction} of star-forming galaxies among the total population
(perhaps switching and quenching star formation).

%----------------------------------------------------------------------
% 4.2 The influence of interaction and/or merger 
%----------------------------------------------------------------------
\subsection{Role of interactions and/or mergers}

In Paper {\sc i}, we found that the fraction of galaxies with a
close companion increases with increasing local galaxy density.  Since
galaxy interactions and mergers can trigger starbursts in galaxies
(e.g., Mihos \& Hernquist 1996; Taniguchi \& Wada 1996), we suggested
that the high fraction of \OII\ emitters in high-density regions may
be driven by interactions and/or mergers.  On the other hand, in the
previous section, we found that the \OII\ fraction in the high-density
environment is higher only for high mass ($M_* \sim 10^{10}-10^{11}
M_{\sun}$) galaxies, while there is no evidence for environmental
dependence at $M_* \lesssim 10^{10} M_{\sun}$.  Here we study the
effects of the interactions and/or mergers on the star formation
activity and its environmental dependence as a function of stellar
mass.

Following Paper {\sc i}, we define a galaxy with a close companion as
a system with a nearest neighbor within an projected separation of
less than 10 arcsec, corresponding to $< 80$ kpc in proper distance at
$z=1.2$.  We use galaxies with $i^{\prime}<24$ within a redshift slice
of $\pm \sigma_{\Delta z}$ to search for nearest neighbors, as in the
estimate of the local galaxy density (Section 3.1).  This redshift
slice corresponds to a maximal velocity difference of $\Delta V \sim
1600$ km s$^{-1}$.  Using this definition, we identify 222 galaxies
with a close companion from the 1446 photo-$z$ selected galaxies.  The
other galaxies do not have a close companion.  We call hereafter these
galaxies ``non-companion galaxies''.

The 71\% of the galaxies with a close companion are \OII\ emitters.
The numbers of companion galaxies for each environment and stellar
mass are summarized in Table \ref{tab:companion}.  Note that these
galaxies may not always be interacting, since the maximum velocity
difference of $\Delta V \sim 1600$ km s$^{-1}$ is large for
interactions or/and mergers.  We calculate the probability of chance
alignment from the average number density of galaxies and the search
area with a radius of 80kpc.  The probability of chance alignment is
0.05 in the full sample and 0.02, 0.05, and 0.12 for the low-,
intermediate-, high-density environments, respectively.  Although the
probability of chance alignment is relatively small, we take this
effect into account in the following analysis.

%--------------------
% table sample (companions)
%---------
%\clearpage
\begin{deluxetable*}{rcccrcccrccc}
%\rotate
\tablecolumns{12}
\tablewidth{0pc}
\tablecaption{\label{tab:companion} The number of companions in each 
environments and mass bins}
\tablehead{
\colhead{}   & \multicolumn{3}{c}{All samples} &
\colhead{}   & \multicolumn{3}{c}{[O {\sc ii}] emitters} &  
\colhead{}   &\multicolumn{3}{c}{non-[O {\sc ii}] emitters} \\
\cline{2-4} \cline{6-8} \cline{10-12}\\
\colhead{$\log M_* (M_{\sun})$} & 
\colhead{Low}   & \colhead{Intermediate}  & \colhead{High} & \colhead{} & 
\colhead{Low}   & \colhead{Intermediate}  & \colhead{High} & \colhead{} &
\colhead{Low}   & \colhead{Intermediate}  & \colhead{High}} 

\startdata
8 -- 8.5 & 0 & 0 & 0 &
         & 0 & 0 & 0 &
         & 0 & 0 & 0 \\
8.5 -- 9 & 2 & 1 & 3 &
         & 1 & 1 & 3 &
         & 1 & 0 & 0 \\
9 -- 9.5 & 19 & 24 & 4 &
         & 17 & 23 & 4 &
         & 2 & 1 & 0 \\
9.5 -- 10 & 21 & 56 & 13 &
          & 17 & 47 & 11  &
          & 4 & 9 & 2 \\
10 -- 10.5 & 8 & 28 & 14&
           & 6 & 13 & 8 &
           & 2 & 15 & 6 \\
10.5 -- 11 & 5 & 15 & 3 &
           & 1 & 2 & 2 &
           & 4 & 13 & 1 \\
11 -- 11.5 & 2 & 4 & 0 &
           & 0 & 2 & 0 &
           & 2 & 2 & 0 \\
11.5 -- 12 & 0 & 0 & 0 &
           & 0 & 0 & 0 &
           & 0 & 0 & 0 \\
\hline
Total &57 & 128& 37 &
      &42 &88 & 28 &
      &15 &40 &9 \\
\enddata
\end{deluxetable*}

%----------------------------------------------------
% fig: emitter fraction vs. M* (companion & noncompanion)
%----------------------------------------------------
%\clearpage
\begin{figure}
\begin{center}
	\includegraphics[width=80mm]{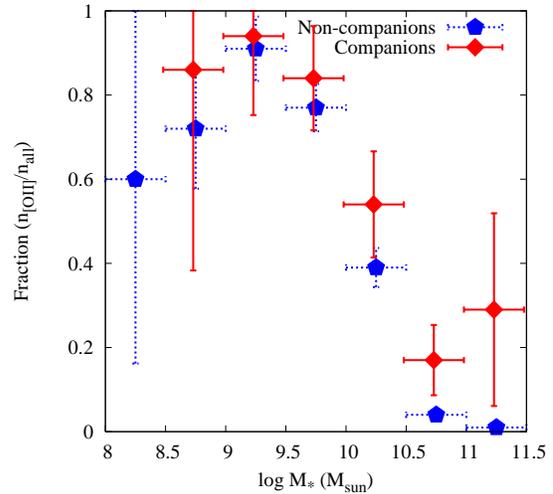} 
\caption{
The fraction of \OII\ emitters as a function of stellar mass
for the galaxies with and without a companion. 
The blue pentagons and red diamonds represent
the without companion and with companion samples, respectively. 
The error bars are the same as Figure \ref{fracoii-mass-env}. 
Data points for the non-companion and companion samples are plotted with small
horizontal offsets for clarity.}
\label{fracoii-mass-compnoncomp}
\end{center}
\end{figure}

%---------------------------------------------
% fig: companion fraction vs. M* (oii & non-oii)
%---------------------------------------------
%\clearpage
\begin{figure}
\begin{center}
	\includegraphics[width=80mm]{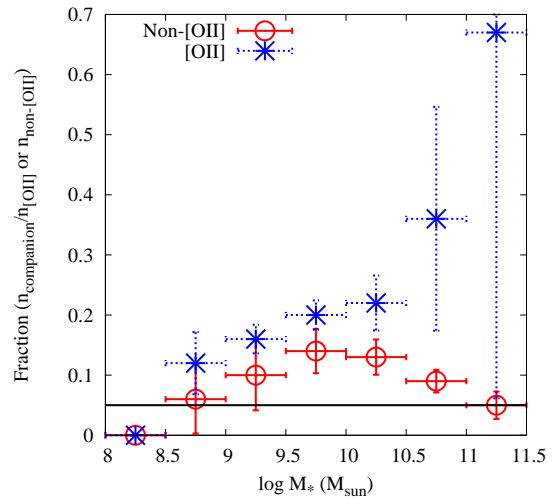} 
\caption{
The fraction of the galaxies with a companion as a function of stellar mass.
The cross and circle points represent \OII\ emitters and
non-\OII\ emitters, respectively. 
The error bars are the same as Figure \ref{fracoii-mass-env}. 
The horizontal line shows the probability of the chance alignment
calculated from average number density of all photo-$z$ selected galaxies.
}
\label{fraccomp-mass-oiinonoii}
\end{center}
\end{figure}

In Figure \ref{fracoii-mass-compnoncomp}, we show the \OII\ fraction
as a function of stellar mass for the galaxies with and without a
companion.  While the \OII\ fractions for these two samples are
similar within the error bars at low mass ($M_* < 10^{10} M_{\sun}$),
the \OII\ fraction for the ``with companion'' sample is $2.4\pm 0.5$
times higher than that for the ``without companion'' sample at $M_* =
10^{10}-10^{11.5} M_{\sun}$.  The difference in \OII\ emitter
fractions between the two samples appears to increase with increasing
stellar mass.  Figure \ref{fraccomp-mass-oiinonoii} shows the relation
between the fraction of galaxies with a companion (hereafter
``companion fraction'') and stellar mass for \OII\ and
non-\OII\ emitters in all the environments.  The horizontal line
represents the expected contribution of the chance alignment.  The
companion fraction for \OII\ emitters is higher than that for
non-\OII\ emitters and the probability of the chance alignment.  It is
also seen that the companion fraction for \OII\ emitters increases
with increasing stellar mass, while the fraction for
non-\OII\ emitters does not.  
Indeed, it is estimated that the companion fraction for \OII\ emitters
at high mass ($M_*=10^{10}-10^{11.5} M_{\sun}$) is $1.5 \pm 0.3$ times higher 
than that at low mass ($M_* < 10^{10} M_{\sun}$) and that for 
non-\OII\ emitters at high mass is only $0.7\pm 0.2$ times that 
at low mass. Moreover, at high mass, the companion fraction for \OII\ 
emitters is $2.8\pm0.6$ times higher than that for non-\OII\ emitters, 
while that for \OII\ emitters is only $1.3 \pm 0.3$ times that for non-\OII\ 
emitters at low mass.
These results suggest that the star formation
activity may be induced by interactions and/or mergers more
preferentially in high-mass galaxies.

\begin{figure}
\begin{center}
	\includegraphics[width=80mm]{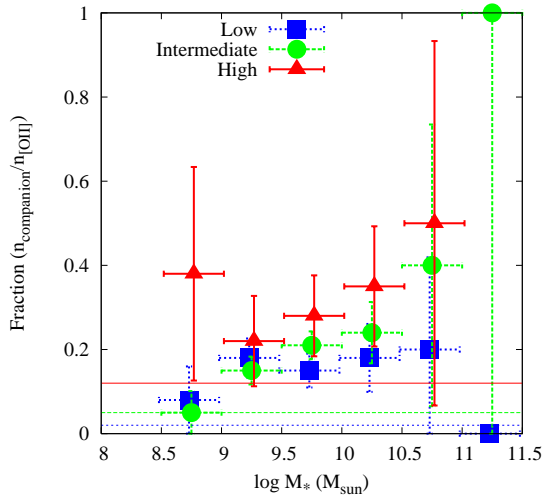} 
\caption{
The fraction of galaxies with a companion 
in \OII\ emitters as a function of stellar mass in
the three different environments.  
The symbols and error bars are the same as Figure \ref{fracoii-mass-env}.
The horizontal blue, green, and red lines show the probabilities of chance alignment
calculated from the average number density of galaxies in the low-, intermediate-, 
and high-density environment, respectively.
Data points for the different environments are plotted with small
horizontal offsets for clarity.
}
\label{fraccomp-mass-env}
\end{center}
\end{figure}

In order to understand the relation between this result and the
environmental dependence of the star formation activity, we also
investigate the companion fraction in \OII\ emitters as a function of
stellar mass in the three different environments (Figure
\ref{fraccomp-mass-env}).  The horizontal lines in the figure
represent the probability of the chance alignment for the three
environments.  
It is seen that the companion fraction tends to be
higher in the higher-density environment, especially at high stellar
mass, although there is the large uncertainty due to the
small number statistics.
The companion fractions at $M_{*}=10^{10}-10^{11.5} M_{\sun}$ 
in the low-, intermediate-, and high-density environment are estimated 
to be $0.2\pm0.1$, $0.3\pm0.1$, and $0.4\pm0.1$, respectively.
Although the probability of the chance alignment also tends to
be higher in the higher-density environment, the environmental
dependence of the companion fraction at high mass ($M_* =
10^{10}-10^{11.5} M_{\sun}$) appears to remain even if we subtract the
contribution of chance alignment.  We also find that the companion
fraction for \OII\ emitters in the high- and intermediate environments
increases with increasing stellar mass.  
The plots of Figure \ref{fraccomp-mass-env}
are fit with the following relations: 
${\rm Fraction (n_{companion}/n_{[OII]})}=(0.03\pm0.03)\times\log M_{*}-(0.16\pm0.34)$
for the low-density environment,
$(0.14\pm0.02)\times\log M_{*}-(1.14\pm0.21)$
for the intermediate-density environment, and
$(0.08\pm0.06)\times\log M_{*}-(0.47\pm0.62)$
for the high-density environment.
Although the uncertainty for
the high-density environment is relatively large due to the small
number statistics, the correlation between the companion fraction and
the stellar mass for the intermediate-density environment is evident.
Thus the star formation in massive \OII\ emitters in higher-density
environment may be preferentially triggered by interactions and/or
mergers.

\begin{figure}
\begin{center}
	\includegraphics[width=80mm]{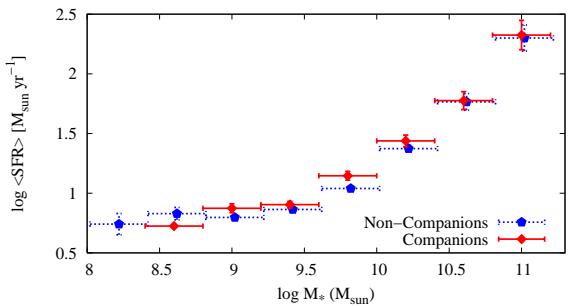} 
\caption{
The average SFR of \OII\ emitters with and without a close companion 
as a function of stellar mass.
The blue pentagons and red diamonds show 
the non-companions and companions, respectively.
The error bars are the same as Figure \ref{sfr-mass-env}.
Data points for the non-companion and companion samples are plotted with small
horizontal offsets for clarity.
}
\label{sfr-mass-compnoncomp}
\end{center}
\end{figure}

Figure \ref{sfr-mass-compnoncomp} shows the average SFR of
\OII\ emitters with and without a companion as a function of stellar
mass. 
We then apply a linear fit to obtain the relation for the \OII\
emitters with and without a companion:
$\log {\rm <SFR>}=(0.59\pm0.11)\times\log M_{*}-(4.59\pm1.01)$ and
$\log {\rm <SFR>}=(0.52\pm0.11)\times\log M_{*}-(3.98\pm1.03)$, respectively.
We find that there is no significant difference between the companion and
non-companion samples.  Since the \OII\ fraction for high-mass
galaxies depends on the presence of a close companion as seen above,
these results imply that galaxy interaction does not affect SFR in
star-forming galaxies but can trigger a larger number of high-mass
galaxies to become star forming at all.

%--------------------------------------------
% fig: companion fraction vs. M* (environment)
%--------------------------------------------
%\clearpage

%------------------------------------------------------------------------------
% fig: sfr vs. stellar mass (companion and non-companion)
%------------------------------------------------------------------------------
%\clearpage
%----------------------------------------------------------------------
% 4.3 Effects of the Contamination and Incompleteness in the sample 
%----------------------------------------------------------------------
\subsection{Effects of the Contamination and Incompleteness in the Sample}

Here we consider the effects of sample contamination and
incompleteness on the results shown in the previous two sections.  As
mentioned in Section 2.3, the estimated photo-$z$ contamination and
incompleteness is $\sim 20\%$.  Since it is unlikely that the fraction
of contamination or incompleteness depends strongly on the local
number density of galaxies, it is expected that the contamination and
incompleteness could not significantly change the estimation of local
galaxy density (although contamination could slightly smear out the
density contrast).

The contamination and incompleteness rates are different between the
\OII\ emitter and non-\OII\ emitter samples; the contamination rates
for the \OII\ emitter sample and for the non-\OII\ emitter sample are
estimated as $8\%$ and $35\%$, and the incompleteness rates for the
\OII\ emitter sample and for the non-\OII\ emitter sample are $0\%$
and $49\%$, respectively.  Since these difference may affect our
results, in the following we discuss their possible effects in detail.

First, we consider possible effects on the \OII\ fraction due to
contamination and incompleteness.  From the contamination and
incompleteness rates for the \OII\ emitter and non-\OII\ emitter
samples, we naively expect that the number of \OII\ emitters is
overestimated by $\simeq 9\%$; i.e., $1/(1-0.08) \approx 1.09$.  The
number of all photo-$z$ selected galaxies (824 \OII\ emitters $+$ 622
non-\OII\ emitters) is underestimated by $\simeq 7\%$; i.e.,
$(824+622)/[824\times(1-0.08)+622\times(1-0.35)/0.51] \approx 1.07$.
Thus we may overestimate the overall \OII\ fraction by $\simeq 17\%$;
i.e., $(1+0.09)/(1-0.07) \approx 1.17$.  In order to investigate
effects on the mass and environmental dependence of the
\OII\ fraction, we check the stellar mass and environment of the
contaminant and missed objects with spectroscopic identification.  We
find that those galaxies for the \OII\ emitter sample and for the
non-\OII\ emitter sample have stellar mass ranges of $10^{9} M_{\sun}<
M_*<10^{10} M_{\sun}$ and $10^{10} M_{\sun}< M_*<10^{11} M_{\sun}$,
respectively, and exist in various environments (although they are
rarer in high-density environments).  Thus contamination and
incompleteness affect all masses and environments evenly, and so they
do not significantly affect the mass and environmental dependence of
the \OII\ fraction seen in Figure \ref{fracoii-mass-env}.  Since the
contamination and incompleteness for \OII\ emitters are small and do
not depend on environment, the stellar mass distribution and the
average SFR at a given mass for these galaxies in the different
environments (Figures \ref{mass-his-oii} and \ref{sfr-mass-env}) are
also not significantly affected.

Next, we consider effects on the results regarding companions.  Since
the incompleteness is expected to be independent of the presence or
absence of a close companion, the missed objects do not affect the
companion fraction.  Similarly the contamination from different
redshifts is not correlated with the spatial distribution of galaxies
at $z = 1.17-1.20$.  Therefore the probability of the contamination
form chance alignment is expected to be roughly the probability of the
chance alignment calculated in Section 4.2 ($\simeq 5\%$).  When the
companion fraction is much higher than the probability of the chance
alignment, the large contamination could lower the companion fraction.
However, high companion fractions are seen for \OII\ emitters at high
mass in Figure \ref{fraccomp-mass-oiinonoii}.  Since the contamination
for \OII\ emitters is small ($\simeq 8\%$), it does not significantly
affect the companion fraction.  On the other hand, the contamination
rate is $\simeq 35\%$ for the non-\OII\ emitters, but the companion
fraction for those galaxies is not much higher than the probability of
the chance alignment (Figure \ref{fraccomp-mass-oiinonoii}).  Then the
effect of the contamination on the companion fraction in the
non-\OII\ emitter sample also seems to be insignificant, although the
companion fraction could be slightly underestimated.

We thus conclude that the contamination and incompleteness of the
photo-$z$ sample do not affect significantly our results.

%----------------------------------------------------------------------
% 5 Discussion
%----------------------------------------------------------------------
\section{DISCUSSION }

In this paper, we have investigated the relations between the star
formation, stellar mass, and environment at $z\simeq 1.2$ in the
COSMOS field.  As shown in section 4, our results are different from
the observational properties of galaxies in the local universe.  Based
on our observational results, we discuss possible origins of the
difference between $z\sim 0$ and $z\sim 1$ from a viewpoint of the
evolution of galaxies.

\subsection{The Observational Properties of Galaxies at $z \simeq 1.2$ in the COSMOS Field}

The findings of this study are summarized by the following:

\begin{enumerate}

\item At $M_* \gtrsim 10^{10} M_{\sun}$, the fraction of \OII\ emitters 
  in high-density environments is $1.7\pm0.4$ times as high as in
  low- and intermediate-density environments.  The fraction of
  \OII\ emitters does not depend on environment at $M_* \lesssim
  10^{10} M_{\sun}$ (Figure \ref{fracoii-mass-env}).

\item The fraction of \OII\ emitters in galaxies with a companion
  (likely interacting galaxies) is $2.4\pm0.5$ times higher than that in
  those without a companion (likely isolated galaxies) over $M_* \sim
  10^{10}-10^{11.5} M_{\sun}$ (Figure
  \ref{fracoii-mass-compnoncomp}).

%\item The fraction of galaxies with a companion for the \OII\ emitters
%  increases with stellar mass, while the fraction of
%  non-\OII\ emitters decreases with stellar mass at $M_* \gtrsim
%  10^{10} M_{\sun}$ (Figure \ref{fraccomp-mass-oiinonoii}).
 
\item The fraction of galaxies with a companion for the \OII\ emitters
  at high mass ($M_{*}=10^{10}-10^{11.5}M_{\sun}$) is $1.5\pm0.3$ times 
  higher than that at low mass ($M_{*}<10^{10}M_{\sun}$), while that for
  the non-\OII\ emitters at high mass is $0.7\pm0.2$ times that at 
  low mass (Figure \ref{fraccomp-mass-oiinonoii}). 

%\item The fraction of \OII\ emitting galaxies with a companion is
%  higher in the higher-density environments and increases with stellar
%  mass in the high- and intermediate-density environments (Figure
%  \ref{fraccomp-mass-env}).

\item The fraction of \OII\ emitters with a companion is higher in the
  higher-density environments (Figure \ref{fraccomp-mass-env}).
  The fractions in the low-,
  intermediate-, and high-density environment at high mass 
  ($M_{*}=10^{10}-10^{11.5}M_{\sun}$) are
  $0.2\pm0.1$, $0.3\pm0.1$, and $0.4\pm0.1$, respectively.

\item The average SFR of \OII\ emitters strongly correlates with
  stellar mass in all the environments. For example, the linear fit 
  gives the relation in the high-density environment,
  $\log {\rm <SFR>}=(0.59\pm0.08)\times\log M_{*}-(4.56\pm0.78)$. 
  The SFR at a given mass is
  independent of the environment and the presence of a close companion
  (Figures \ref{sfr-mass-env} and \ref{sfr-mass-compnoncomp}).

\end{enumerate}

From the first item described above, it is expected that high-mass
\OII\ emitters contribute significantly to the active star formation
in high-density regions at $z\simeq 1.2$.  Moreover, items 2, 3, and 4
suggest that interactions and/or mergers could trigger star formation
in massive galaxies preferentially in high-density environments, while
the star formation in low-mass galaxies appears to be independent of
environment and does not seem to be affected by the interactions
and/or mergers.  In other words, $EW$(\OII) of some inactive high-mass
galaxies may become higher due to the active star formation induced by
galaxy interactions, and thus we observe such high-mass galaxies as
\OII\ emitters.  On the other hand, considering item 5, it appears
that the strength of star formation activity in the star-forming
galaxies is not influenced by environmental effects, although the
environment and interactions affect the fraction of star-forming
galaxies.

The environmental dependence of the \OII\ fraction is seen only at
high mass.  It is interesting to note that the influence of
interactions and/or mergers on star formation depends both on
environment and stellar mass.  This can be interpreted such that
massive galaxies need external triggers (i.e., interactions and/or
mergers) for active star formation, while the less massive galaxies
are form stars regardless of the presence of a close companion.  One
possible triggering mechanism for less massive galaxies is minor
merger events (e.g., Taniguchi \& Wada 1996).  However, such satellite
galaxies would be too faint to be detected in our imaging survey.  As
an another possible interpretation, it is expected that less massive
galaxies contain much gas and can naturally form stars since evolution
of those galaxies is slow compared to massive galaxies, which is
expected from mass-downsizing (e.g., Cowie et al. 1996).

Therefore, we conclude that the environmental dependence of the
\OII\ fraction (i.e., the fraction of star-forming galaxies) is seen
only at high mass, because 1) interactions and/or mergers induce the
star formation only in massive galaxies and 2) the probability of
interactions and/or mergers depends on local galaxy density.

Peng et al. (2010, 2011) recently found that the environment quenching
(or the satellite quenching) depends on the local galaxy density but
not on the stellar mass of galaxies.  We observe a different
environmental effect from that found by Peng et al. (2010, 2011),
since our results suggest that the influence of interaction and/or
mergers depend on both environment and stellar mass.

\subsection{Implications to the evolution of galaxies from $z\sim 1$ to $z\sim 0$}

In the local universe, it is observed that the fraction of red
galaxies in high-density environments is higher than that in
low-density environments, and this difference becomes larger at lower
mass (e.g., Baldry et al. 2006).  Namely, the fraction of blue
galaxies in high-density environments is lower than that in
low-density environments, especially at low mass.  Iovino et al (2010)
also found a similar trend at $z<0.5$.  Furthermore, they found that
the fraction of blue galaxies at $z=0.6-0.8$ does not depend on the
environment in $M_* > 10^{10}$.  These are in contrast to the
environmental dependence of the \OII\ fraction in massive galaxies at
$z\simeq 1.2$ mentioned above.

Now a question arises: what is the origin of the difference between
$z\sim 1$ and $z\sim 0$?  Our results suggest that interactions and
mergers are likely to induce star formation in massive galaxies in
high-density environments at $z\simeq 1.2$, resulting in the observed
high \OII\ fraction of massive galaxies in high-density environments.
These inactive massive galaxies where star formation would be induced
by interactions are likely to have some cold gas for star formation.
We have also found that the average SFR at a given mass is similar
among the different environments and is independent of the presence of
a close companion (Figures \ref{sfr-mass-env} and
\ref{sfr-mass-compnoncomp}), the strength of the star formation
induced by interactions in massive galaxies does not seem to be
significantly different from other massive star-forming galaxies on
average.  Therefore we expect that there are already some inactive
high-mass galaxies with sufficient cold gas for SFR at a given mass at
$z\simeq 1.2$.  On the other hand, the fraction of gas-poor galaxies
is higher in higher density regions at low-$z$ (e.g., Giovanelli \&
Haynes 1985; Bertram et al. 2006).  Therefore it is unlikely that a
significant starburst occurs in interactions or mergers between
gas-poor galaxies (i.e., dry mergers; e.g., Tran et al. 2005; Cattaneo
et al. 2008), even though interactions and/or mergers still occur in
high-density regions at low-$z$ (e.g., Patton et al. 2011).

In this context, one important difference in galaxies between $z\sim
1.2$ and low $z$ is the cold gaseous content.  Since galaxies at
$z\sim 1.2$ are in early stages of their evolution, it is likely that
their cold gaseous content is, on average, larger than that in
galaxies at low $z$ even if such galaxies are located in relatively
high-density regions.  It is thus expected that interactions and
mergers at $z\simeq 1.2$ could cause intense star formation more
frequently since it is known that interactions and mergers between two
gas-rich galaxies (wet mergers) can induce active star formation and
rapidly consume the cold gas (e.g., Barton et al. 2000; Woods et
al. 2006).  In fact, Lin et al. (2008, 2010) suggest that the fraction
of dry mergers gradually increases from $z\sim 1$ to $z\sim 0$, while
the fraction of wet mergers decreases at the redshift range.

If there are many inactive high-mass galaxies with sufficient cold gas
for star formation at $z\simeq 1.2$ as mentioned above, there may be
some mechanisms responsible for quenching star formation in galaxies
while retaining their gas.  For example, it might be expected that
high-mass galaxies have experienced starbursts in the past, and then
their gas becomes hot and their star formation stops in spite of the
existence of the (hot) gas.  Later the hot gas cools down into cold
gas, and then high-mass galaxies actively form stars when interactions
and/or mergers occur.  As another scenario, high-mass galaxies might
have consumed most of gas in the past, which is expected from the
mass-downsizing of galaxy formation (e.g., Cowie et al. 1996), and the
gas density was already too low to cause intense star formation
(Kennicutt 1989).  However, if interactions can sufficiently perturb
the remaining gas, the star formation can be triggered and they can be
observed as high-mass star-forming galaxies in high-density
environment.  In this case, such galaxies immediately consume their
gas in the star formation and these massive galaxies in high-density
environment are observed to be passive at $z \lesssim 1$.

From these considerations, we suggest the following scenario; The
high-mass star-forming galaxies for which star formation are induced
by interactions and/or mergers contribute significantly to the active
star formation in high-density regions at $z\simeq 1.2$.  These
massive star-forming galaxies in high-density regions at $z\simeq 1.2$
could quickly consume most of the accreted cold gas.  If this is the
case, the star formation activity may not be enhanced when
interactions and/or mergers occurred in high-density environments at
lower redshift.  In this context, the quenching of star formation in
massive galaxies in high density environments is expected to lead to
the shift of major star formation in the universe from high-density
regions to low-density ones at $z\lesssim 1$.

We would like to thank the anonymous referee for her/his very useful
comments and suggestions.  We also thank all members of the COSMOS team.  
This work was financially supported in part by the Japan Society for the 
Promotion of Science (Nos. 17253001, 19340046, 23244031, and 23654068).
Y. I. is financially supported by the Japan Society for the Promotion
of science (JSPS) through JSPS Research Fellowship for Young
Scientists.

%------------------------------------------------------------------------------
%    References
%------------------------------------------------------------------------------

\end{document}